\documentclass[aps,prd,amsmath,preprintnumbers]{revtex4}

\usepackage{epsfig}
\usepackage{graphicx}

\renewcommand{\vec}[1]{\mathbf{#1}}  
\newcommand{\mq}{|\vec{q}|}
\newcommand{\fm}{\text{fm}}
\newcommand{\MeV}{\text{MeV}}

\newcommand{\GeV}{\text{GeV}}

\newcommand{\tmu}{\tilde{\mu}}

\newcommand{\be}{\begin{equation}}
\newcommand{\ee}{\end{equation}}
\newcommand{\bea}{\begin{eqnarray}}
\newcommand{\eea}{\end{eqnarray}}
\newcommand{\bean}{\begin{eqnarray}}
\newcommand{\eean}{\end{eqnarray*}}

\newcommand{\gapproxeq}{\lower
.7ex\hbox{$\;\stackrel{\textstyle >}{\sim}\;$}}
\newcommand{\lapproxeq}{\lower
.7ex\hbox{$\;\stackrel{\textstyle <}{\sim}\;$}}

\def\3bar{$\bar {\hbox{\bf 3}}$}

\newcommand{\cc}{$c\bar{c}$}

\renewcommand{\bar}[1]{\ensuremath{\overline{#1}}}

\begin{document}

\title{Novel Charmonium and Bottomonium Spectroscopies due to Deeply Bound Hadronic Molecules from Single Pion Exchange}
 
\author{Frank Close}
\email[E-mail: ]{f.close1@physics.ox.ac.uk}
\affiliation{Rudolf Peierls Centre for Theoretical Physics, University of Oxford,
\\ 1 Keble Road, Oxford, OX1 3NP, UK}
 
\author{Clark Downum}
\email[E-mail: ]{c.downum1@physics.ox.ac.uk}
\affiliation{Clarendon Laboratory, University of Oxford,\\ Parks Road, Oxford, OX1 3PU, UK}

\author{Christopher E. Thomas}
\email[E-mail: ]{thomasc@jlab.org}
\affiliation{Jefferson Laboratory, 12000 Jefferson Avenue, Suite \#1,
\\ Newport News, VA. 23606 USA}

\preprint{OUTP-10-01P}
\preprint{JLAB-THY-10-1119}

\begin{abstract}
Pion exchange in S-wave between hadrons that are themselves in a relative S-wave is shown to shift energies by hundreds of MeV, 
leading to deeply bound quasi-molecular states. 
In the case of charmed mesons $D^*,D_1$ a spectroscopy arises
consistent with enigmatic charmonium states observed above 4 GeV in $e^+e^-$ annihilation.
A possible explanation of $Y(4260)\to \psi\pi\pi$ and $Y(4360) \to \psi'\pi\pi$ is found.
We give results for all isospin and charge-conjugation combinations, and comment on flavor exotic doubly charmed states and bottomonium analogs.
A search in $D\bar{D}3\pi$ 
is recommended to test this hypothesis. 
An exotic $1^{-+}$ is predicted
to occur in the vicinity of the $Y$(4260).
\end{abstract}

\maketitle

\section{Introduction}
\label{sec:Introduction}

If two hadrons $A,B$ are linked by $A \to B \pi$, then necessarily hadronic pairs of $AB$ or 
$A\bar{B}$ have
the potential to feel a force from $\pi$ exchange. This force will be attractive in
at least some channels. Long ago\cite{torn,ericson} the idea of $\pi$ exchange between flavored mesons, in 
particular charm,
was suggested as a source of potential ``deusons"\cite{torn}. Using the deuteron binding as 
normalization, the
 attractive force between the $J^P=0^-$ charmed $D$ and its $J^P=1^-$ counterpart $D^*$ was calculated 
for the
 $D\bar D^* + c.c.$ S-wave combination with total $J^{PC}=1^{++}$, and the results compared with the 
enigmatic
 charmonium state $X(3872)$\cite{pdg08,torn,classics,fcthomas}.
 
 In all these examples, as in the more traditional case of the nucleon, where the $NN\pi$ coupling is the source of an 
attractive force that helps to
form the deuteron, the exchanged $\pi$ was emitted and absorbed in a relative P-wave with respect to the hadrons. 
In such cases, the binding energies that result are ${\cal O}(1-10)$MeV; this
in particular has encouraged interest in the $X(3872)$ which is within errors degenerate with the $D^0D^{*0}$ threshold.
It has recently been pointed out\cite{cdprl} that 
the exchange of a $\pi$ in S-wave, between pairs of hadrons that are themselves in relative S-wave,
leads to deeply bound states between those hadrons. Instead of binding of a few MeV, as in the cases considered historically,
there is now the potential for binding on the scale of ${\cal O}(100)$MeV, leading to a rich spectroscopy of states that are far from the di-hadron
channels that create them. We shall argue that examples of such a spectroscopy appear to
be manifested among charmonium-like mesons. 

We organize this paper as follows.
First we summarize the general arguments for expecting large binding energies due to S-wave $\pi$ exchange. 
We shall consider a chiral Lagrangian model to illustrate and quantify the phenomenon of energy shifts of 
$\mathcal{O}(100)$MeV, focusing specifically on the charmonium-like $1^{--}$ isoscalar ($I=0$) channel.  
In Section \ref{sec:ChiralModel} we investigate the connection between the chiral
potential and the decay width of the relevant charmed mesons, first in the heavy quark limit with point particles
and subsequently in the non-heavy quark limit and with form-factors.
Then, we solve the Schr\"odinger equation and discuss the uncertainties within the model.  Detailed results
for the charmonium-like $1^{--}, I=0$ are presented in Section \ref{sec:Spectroscopy} along 
with results for other $J^{PC}$, isospin, and flavor channels.
We discuss the limitations of our approximation to the strong interaction in Section \ref{sec:Uncertaintities}, 
give phenomenological implications and suggest experimental searches in 
Section \ref{sec:Phenomenlogy} and finish with conclusions in Section \ref{sec:Conclusions}.

\section{Molecules and S-wave $\pi$ exchange} 
\label{sec:GeneralArguments}
Several groups have studied 
the following meson-pairs looking for bound states in total $J^{PC}$ channels due to pion exchange
(from here on $D\bar D^*$ etc. will be taken to include the charge conjugate channel):
\begin{eqnarray*}
D^*(1^-) \to D(0^-) \pi &\textrm{ leading to the deuson } \bar{D}D^*\; X(3872)\ J^{PC}=1^{++}\\
D^*(1^-) \to D^*(1^-) \pi &\textrm{ leading to the deusons } D^*\bar D^* \; J^{PC}=0^{++},1^{+-},2^{++}
\end{eqnarray*}
 These combinations were discussed in \cite{torn}. In all of these examples parity conservation 
requires that 
the $\pi$ is emitted in a P-wave; the hadrons involved at the emission vertices
 have their constituents in a relative $s$-wave, (we use S,P to denote the angular momentum between
 hadrons, and $s,p$ to denote internal angular momentum of the constituents within a hadron). 
 In such cases, the $\pi$ emission being in P-wave causes a penalty for small momentum
 transfer, $\vec q$, which is manifested by the interaction\cite{ericsonwise}
 
\begin{equation}
V_P(\vec{q}) = - \frac{g^2}{f_{\pi}^2} \frac{(\vec{\sigma}_i \cdot \vec{q})(\vec{\sigma}_j \cdot 
\vec{q})}
{\mq^2 + \mu^2}(\vec{\tau}_i \cdot \vec{\tau}_j)  .
\label{pwavepi}
\end{equation}
\noindent where $\mu^2 \equiv m_{\pi}^2 - (m_B-m_A)^2$, $m_{A,B}$ being the masses of the mesons in 
the process $A \to B\pi$.
(For a discussion of this interaction, and its sign, see Eq.\  (20) in ref\cite{fcthomas}).
\noindent The resulting potential is $\propto \vec{q}^2$ for low momentum transfer and has been found 
to give bindings on the scale of a few MeV, which is in part
a reflection of the P-wave penalty.

There is no such penalty when $\pi$ emission is in S-wave, which is allowed when the hadrons $A,B$
have opposite parities. Examples 
involving
the lightest charmed mesons are $D_1(1^+) \to D^*(1^-) \pi$ and $D_0(0^+) \to D(0^-) \pi$.
P-wave $\pi$ exchange carries a $\vec{q}$ penalty at each $\pi A B$ vertex.
One might naively anticipate that the transition from a $D$ or $D^*$, with constituents in $s$-wave, to the $D_1$ or $D_0$, 
with constituents in $p$-wave, would restore a $\vec q$ penalty, leading to small binding 
effects as in the
cases previously considered. However, as we now argue, this need not be the case, and energy shifts of 
$\mathcal{O}(100)$MeV can arise.

There is a long history of data on $\pi$ transitions between hadrons of opposite parity
which indicate that the S-wave coupling is significant when $\vec{q} \to 0$.
In the charm sector of interest here, the large widths\cite{pdg08} for $\Gamma(D_0 \to D\pi) \sim 260 
\pm 50$MeV
 and $\Gamma(D_1(2430) \to D^*\pi) \sim 385 \pm \mathcal{O}(100)$MeV suggest that, even after phase space is 
taken into account,
there is a significant transition amplitude. This non-suppression 
was specifically commented upon in the classic quark model paper of 
ref.\cite{fkr}. It arises from a derivative operator acting on the internal
hadron wave function, which enables an internal $s$ to $p$ transition to occur without suppression
even when $\vec q$ vanishes. 
This can be seen when $\bar{\psi}\gamma_5\psi$ is expanded to $\sigma.(\vec{q}-\omega 
\vec{p}/m)$, where $\vec{p}$ the internal quark momentum\cite{divgi}.
 Feynman, {\it et al.}\cite{fkr} argued for this form on general grounds of Galilean invariance.
 The presence of $\vec{p}$ gives the required derivative operator, and hence the
 unsuppressed $p \to s$ transition.

 An unsuppressed transition, when applied to $\pi$ exchange in the $D_1\overline{D^*}$ system 
(e.g. \cite{liu}) causes the chiral model analogue of 
Eq.\eqref{pwavepi} in the $I=0, 1^{--}$ channel to become
  \begin{equation}
    V_S(\vec{q}) =\frac{h^2}{2f_\pi^2}\frac{(m_{D_1} - m_{D^*})^2}
{\mq^2 + \mu^2}
(\vec{\tau}_i \cdot \vec{\tau}_j)  {\cal F}(\vec q)^2
\label{swavepi}
\end{equation}
where $h/(\sqrt{2}f_{\pi})$ is the $D_1D^*\pi$ coupling constant (up to a phase), $f_\pi=132$ MeV, 
$\vec{q}$ is the exchanged three-momentum, $\mu^2 \equiv -(m_{D_1}-m_{D^*})^2 + m_\pi^2$ 
($\mu^2<0$ for the $D_1D^*$ system),  
and $(\vec{\tau}_i \cdot \vec{\tau}_j)$ is the usual contraction of Pauli matrices resulting from
the exchange of an isovector by two isospin-half particles.  ${\cal F}$ is the model dependent 
form-factor which regulates the potential and would be unity in the chiral model.

In the derivation of the potential a static approximation has been made to the pion propagator.
The full propagator is $q^2 - m_\pi^2 = (E_A - E_B)^2 - \vec q^2 -m_\pi^2$.  Approximating 
$E_A =m_A$ and $E_B = m_B$ one recovers the form of the propagator presented in the potential, 
Eq.~\eqref{swavepi}.

The potential is similiar to one presented in Table 1 of Ref.\ \cite{liu}, who were investigating a
$\pi$ exchange model of the $1^{-}, I=1$ $Z^+$(4430).  They considered only the $I=1$ channel and 
omitted the $\vec{\tau}_i \cdot \vec{\tau}_j$ factor (which is unity for $I=1$).
We have made this factor explicit as it will become crucial when we study the $I=0$ sector later.
The absence of a $\vec{q}^2$ penalty factor is immediately apparent. The scale is now being set by
the mass gap squared, which is equal to the timelike component of the momentum transfer vector squared, 
$q_0^2\rightarrow(m_{D_1}-m_{D*})^2$ as $\mq \rightarrow 0$.

This potential and any bound states have immediate implications for a rather rich set of physics.
The potential also applies for the 
$D_0\overline{D}$ system, and the bottomonium and strangeonium
 analogs of $D_1\overline{D^*}$ and $D_0\overline{D}$,
by exchanging the masses with their appropriate counterparts.  
Note that the potential in Eq.~\eqref{swavepi} has no spin dependence and therefore any results apply
equally to the $D_1 \overline{D^*}$ spins coupled to total spin $0$, $1$ or $2$.  For example, if an isoscalar $1^{--}$ 
bound state is found, then we also expect degenerate $0^{--}$ and $2^{--}$ states.
  
Thus on rather general grounds we may anticipate significant energy shifts, 
$\sim \mathcal{O}(100$MeV), due to $\pi$ exchange at least in some channels 
between such hadrons in a relative S-wave.  Signals may be anticipated below 
or near threshold in the following channels (in the charmonium analogues, 
involving charm and anti-charm
mesons for either $I =$ 0 or 1, or in exotic states with manifest charm
involving two charm mesons):
\begin{eqnarray*}
  D_0(0^+) \to D(0^-) \pi &\textrm{leading to the deusons }\phantom{1}\, D\bar D_0\; 
J^{PC}=0^{-\pm}\phantom{12345}\\
  D_1(1^+) \to D^*(1^-) \pi &\textrm{leading to the deusons } D^*\bar D_1 \; J^{PC}=(0,1,2)^{-\pm}\\
\end{eqnarray*}
We also find that it is possible that $L>0$ states could bind which would lead to more $J^P$
configurations.

Pion exchange depends on the presence of $u,d$ flavors, therefore  
there will be no such effects in the $D_s\overline{D}_s$ analogues.
Further, the potential depends only on the quantum numbers of the light quarks.  
Therefore, there will be effects in the strange and bottom 
analogues, which can add to the test of such dynamics at different kinematics.

The parameter $h$ in Eq.~\eqref{swavepi} is closely connected to the width of the $D_1\to D^*\pi$ decay.
Data exists on this decay which constrains the value of $h$ and hence the spectrum of the model.
We discuss the extraction of $h$ from the decay width in the next section.

\section{The Coupling Constant $h$ }
\label{sec:ChiralModel}


Being simply related to the $D_1D^*\pi$ coupling constant, $h$ also appears in the chiral 
formula for the $D_1 \to D^*$ decay width.  Eq.~(137) of Ref.\ \cite{casalbuoni97} gives:
\begin{equation}
 \Gamma(D_1^0 \to D^{*+} \pi^-) = \frac{1}{2\pi}\left(\frac{h}{f_{\pi}}\right)^2 (m_{D_{1}} - m_{D^{*}})^3.
\label{d1width}
\end{equation}
which is valid in the heavy quark limit.  

\begin{table}
  \begin{tabular}{lccccc}
    $A\to B\pi$  & $m_A$/MeV & $m_B$/MeV & $\Gamma$/MeV & BF & $\mq$/MeV\\
    \hline
    $D_1(2430)\to D^*(2010)\pi$ & 2427 $\pm$ 40 & 2010.27 $\pm$ .17 & 384 $^{+130}_{-110}$ & N/A & 359\\
    $D^*_0(2400)^0\to D\pi$ & 2352 $\pm$ 50 & 1896.62 $\pm$ .20 & 261 $\pm$ 50 & N/A & 414\\
    $D^*_0(2400)^\pm \to D\pi$ & 2403 $\pm$ 40 & 1864.84 $\pm$ .17 & 283 $\pm$ 40 & N/A & 461\\
    $B_1(5721)\to B^*(5325)\pi$ & 5723.4 $\pm$ 2.0 & 5325.1 $\pm$ .5 & N/A & dominant & 360 \\
    $K_1(1400)^\pm \to K^*(892)\pi$ & 1403 $\pm$ 7 & 891.66 $\pm$ .26 & 174 $\pm$ 13 & 94 $\pm$ 6\% & 402\\
    $K^*_0(1430)^\pm \to K\pi$ & 1425 $\pm$ 50 & 493.677 $\pm$ .016 & 270 $\pm$ 80 & 93 $\pm$ 10\% & 619\\
  \end{tabular}
  \caption{\label{tab:PDGData} Data of low lying mesons of different flavor sectors with opposite parity 
  which exhibt a large width.
  Values taken from the Particle Data Group\cite{pdg08}.  No width data are available for the bottom sector and 
  no branching fractions are given for the charmed sector.}
\end{table}
In order to extract $h$ using Eq.~\eqref{d1width}, we use the data from the PDG listed in
Table~\ref{tab:PDGData}.  In the absence of a branching fraction we assume that the total 
width is saturated by the $D^*\pi$ channels.  
We are using chiral formulae for the charged $\pi$ width which may be related to the total 
$\pi$ decay width by $\Gamma(D_1^0 \to D^{*+} \pi^-) = \frac{2}{3} \Gamma(D_1^0 \to D^{*} \pi)$
\cite{falkluke}.
Therefore, we use $m_{\pi^+} = 140.$MeV and $f_\pi = 132$MeV throughout.
For the $D_1\to D^*\pi$ system we have $h = 0.63^{+.16}_{-.13}$.

There are theoretical and empirical reasons to suspect that Eq.~\eqref{d1width} may be a poor estimate
for $h$ given $\Gamma$.  Firstly, in the heavy quark limit assumed for Eq.~\eqref{d1width},  
$m_{D_1} = m_{D_0}$, $m_{D^*} = m_D $, and thus $\Gamma ( D_1 \rightarrow D^*\pi)
= \Gamma(D_0\rightarrow D\pi)$ as Eq.~\eqref{d1width} applies equally well to the $D_0\rightarrow D\pi$ 
decay.  However, these relations do not hold experimentally.  Finite mass effects (including 
mass differences) have been used to derive a correction to Eq.~\eqref{d1width}\cite{casalbuoni97}:
\begin{equation}
  \Gamma(D_1^0 \to D^{*+} \pi^-) =
    \frac{h^2}{8\pi f_{\pi}^2} \frac{|\vec{q}| m_{D^*}}{m_{D_1}^3} (m_{D_1}^2-m_{D^*}^2)^2 
    \times \frac{1}{3}\left( 2+\frac{(m_{D_1}+m_{D^*})^2}{4m_{D_1}^2m_{D^*}^2} \right)
\label{d1widthfs}
\end{equation}
Using this expression we have $h = 0.80^{+.20}_{-0.17}$.

We have mentioned that our analysis of the $D_1\overline{D^*}$ system applies equally well to the
$D_0\overline{D}$ system.  Indeed, Eq.~\eqref{d1width} applies to both systems with a trivial substition 
of the appropriate mass.  However, when finite mass effects are included, chiral model 
gives a different formula for the decay widths of the $D_1$ and $D_0$ mesons.
The analogous formula to Eq.~\eqref{d1widthfs} is\cite{casalbuoni97}
\begin{equation}
  \Gamma(D_0 \to D^+ \pi^-) =
    \frac{h^2}{8\pi f_{\pi}^2} \frac{|\vec{q}| m_{D}}{m_{D_0}^3} (m_{D_0}^2-m_{D}^2)^2 .
\label{d0widthfs}
\end{equation}

Due to the larger mass gap (and hence the larger $\mq$), 
Eqs.~\eqref{d1widthfs}  and \eqref{d0widthfs} imply that $\Gamma(D_0\rightarrow D\pi) \approx 1.5\Gamma(D_1\rightarrow D^*\pi)$.
Empirically\cite{pdg08}, 
$$\Gamma(D_0 \to D\pi) \sim 260 \pm 50\ \textrm{MeV and } \Gamma(D_1(2430) \to D^*\pi) \sim 385 \pm \mathcal{O}(100)\ \textrm{MeV}.$$
 Although the uncertainties are large, $\Gamma(D_0 \to D\pi)$ has a smaller width than 
 $\Gamma(D_1 \to D^{*} \pi)$ even though the phase space is larger, in contrast with the expectations
 of Eqs.~\eqref{d1widthfs} and~\eqref{d0widthfs}.

In general, processes such as $\Gamma(D_1^0 \to D^{*+} \pi^-)$ involve form factors that summarize the
penalty for selecting the exclusive process of single $\pi$ emission, which is increasingly improbable 
at large $\mq$ relative to multi-pion, inclusive, channels.  Thus, the assumption that the $D_1D^*\pi$ 
($D_0D\pi$) coupling is constant in the chiral model does not take account of
the full dynamics at the vertex.  The data suggest that we must include the effects of exclusive form factors.

The effects of form factors may be modelled by making the replacement $h \rightarrow h\mathcal{F}(\mq)$ everywhere.
$\mathcal{F}(\mq)$ is a smooth, decreasing function such that $\mathcal{F}(\mq = 0)=1$.  The exact form of
$\mathcal{F}$ is model dependent; however, the introduction of a form factor will in general lead to 
an increased estimate of $h$ and so, naively, to an increased binding energy.

As an explicit example, consider the form factor resulting from a dynamical model of $\pi$ emission\cite{cs}:
\begin{equation}
\mathcal{F}(x) = \left(1-\frac{2}{9}x^2\right)e^{-x^2 / 12}
\label{form-factor}
\end{equation}
with $x \equiv \mq/\beta$ and $\beta \sim 0.4$GeV\cite{cs}.
For $D_1^0 \to D^{*+} \pi^-$ one has $x = 0.89$ while for $D_0^0 \to D^+\pi^-$ $x= 1.18$. 
This plays a significant role in the relative widths as 
\begin{equation}
\Bigl[\frac{\mathcal{F}(x = 0.89)}{\mathcal{F}(x = 1.18)}\Bigr]^2 = 1.6
\end{equation}
which drives the ratio of widths in favour of the $D_1$.  

In turn the form-factor also shows that $h$,
extracted earlier from the chiral model, is an underestimation.
In such a model the more general Eq.~\eqref{d1widthfs}
modified the heavy quark value of $h = 0.63 ^{+.16}_{-.13}$ to $h = 0.80 ^{+.20}_{-.17}$ and the effect of form-factors
further increases $h$ to $h = 1.0^{+0.3}_{-0.2}$.  Therefore, the inclusion of finite mass corrections 
and the effects of exclusive form factors has a significant impact on the value of $h$ extracted from 
experimental decay widths. We emphasise that although the form factor itself is model dependent, the suppression 
for larger $\mq$ is expected in general.

\begin{table}
  \begin{tabular}{lccc}
    System \hspace{.2in}      &  \hspace{.4in} HQ\hspace{.4in}   & \hspace{.0in}  NHQ \hspace{.4in}&\hspace{.3in} NHQFF\hspace{.4in} \\
    \hline
    $D_1(2430)\to D^*(2010)\pi$&  $0.63^{+.16}_{-.13} $ & $0.80^{+.20}_{-0.17}$ & $1.0^{+0.3}_{-0.2}$ \\
    $D^*_0(2400)^0\to D\pi$    &  $0.41\pm0.06$ & $0.55\pm0.08$ & $0.79\pm 0.11$ \\
    $D^*_0(2400)^\pm \to D\pi$ &  $0.36\pm0.04$ & $0.50\pm0.05$ & $0.80\pm 0.08$ \\
    $K_1(1400)^\pm \to K^*(892)\pi$&  $0.30\pm0.02$ & $0.50\pm0.04$ & $0.70\pm 0.05$ \\
    $K^*_0(1430)^\pm \to K\pi$ &  $0.15\pm0.04$ & $0.47\pm0.13$ & $1.2\pm 0.3$ 
  \end{tabular}
  \caption{\label{tab:hvalues}  Values of $h$ extracted for various systems which may experience S-wave
  $\pi$ exchange.  The adaptation of equations for the charm-system to their appropraite form for flavor analog systems
  by making obvious mass substitutions is assumed.  }
\end{table}

In summary, from these different determinations we find values ranging from $h \approx 0.5$ to $1.3$: 
the value of $h$ is highly model and data dependent.  We collect these results and present other results
for analogous systems in Table\ \ref{tab:hvalues}.  
The HQ column presents the values of $h$ extracted in the heavy quark limit using Eq.~\eqref{d1width}.
  The NHQ column is similiar but extracts the values in the non-heavy quark limit using Eqs.~\eqref{d1widthfs} and
  \eqref{d0widthfs}.  The NHQFF column presents extracted $h$ values which would be required to overcome the form-factor
  suppresion, Eq.~\eqref{form-factor}, and to reproduce the correct width in the non-heavy quark limit.  
  In the following section we will present results for 
a range of $h$ and show that the spectrum is highly sensitive to the value of $h$.


\section{Molecular Spectroscopy}
\label{sec:Spectroscopy}

Previously\cite{cdprl} we performed a variational calculation with the potential in Eq.~\eqref{swavepi} 
and $h = 0.8 \pm 0.1$ using trial wave functions.  With this technique we agreed with Ref.\ \cite{liu}
that there was no reason to expect an isovector $1^{--}, D_1 \bar{D}^*$ bound state.  Additionally, we found deep
binding in the isoscalar $D_1 \bar{D}^*$ system.  The presence of deep binding in the 1$^{--}$ channel 
motivated the present study where we solve the Schr\"{o}dinger equation and analyze the 
spectroscopy emerging from S-wave $\pi$ exchange binding of the $D_1 \bar{D}^*$ and analogous systems.  

We solve the Schr\"odinger equation and quantify the bound states from S-wave $\pi$ exchange
using a range of $h$ to set the scale.  The resulting spectrum contains several potential bound states. 
The $1^{--}, I=0$ channel includes 1S and 2S states which are consistent with the $Y(4260)$ and $Y(4360)$ structures 
claimed in $e^+e^-$ annihilation.  
Results for the charmonium-like exotic $1^{-+}$ and isovector $1^{--} $ channels are also presented.
We find the binding energies are highly sensitive both to the value used for $h$ in all channels, 
and attempts to model finite size effects in some channels.
We first consider the potential from the chiral model 
involving a pointlike interaction, and then discuss modification of the potential due to 
finite size effects.

\subsection{Point-like pion exchange}
\label{sec:results:pointlike}

The Fourier transform of Eq.\eqref{swavepi} gives the $D_1 \bar{D}^*$ $1^{--}$ potential 
with S-wave $\pi$ exchange in coordinate space.
When $\mu^2 \equiv -\tmu^2 < 0$ the real part of the potential is: 
\begin{equation}
  V_{\rm S}(r) =\frac{h^2 (m_{D_1}-m_{D^*})^2 }{8 \pi f_{\pi}^2} \frac{\cos(\tmu r)}{r} (\tau_i\cdot\tau_j)
\label{equ:swavepi_posspace}
\end{equation}
 in agreement with Ref. \cite{liu}.  
We numerically solve the Schr\"{o}dinger equation using this position space potential as
described in Ref.\ \cite{fcthomas}. 

Clearly the results for larger values of $h$, which yield significant binding, 
will have important finite size corrections.  Therefore the point particle results can only be considered
to give a cursory quantitative examination of the implications of our general argument.
However, given the unusual nature of the oscillatory potential, it is beneficial to study the
unregulated potential in order to contextualize the effects of a form-factor which are explored
in the next subsection.

In Table \ref{table:isoscalarcharm} we show the binding energies of some of the low-lying isoscalar 
$D_1 \bar{D}^*$ states in relative S-wave with $C=-$ (the parity obviously 
depends on the relative orbital angular momentum of the system; since the potential is independent
of spin, the binding energies are degenerate across all possible total $J$ combinations).  Binding energies 
are given for a few values of $h \in [0.8, 1.3]$.  The binding energies are seen to be very sensitive 
to the value of $h$. 

\begin{table}[htp]
\begin{tabular}{l|c|c|c|c|c|c|}
 & \multicolumn{6}{c}{Binding Energy / MeV} \\
  State           & $h=0.8$ & $h=0.9$ & $h=1.0$ & $h=1.1$ & $h=1.2$ & $h=1.3$ \\
\hline
1S $(0,1,2)^{--}$ & $230$   & $415$   & $680$   & $1000$  & $1500$  & $2100$  \\
2S                & $12$    & $20$    & $29$    & $39$   & $76$    & $210$   \\
3S                & $1.5$   & $3.6$   & $6.7$   & $11$    & $51$    & $65$    \\
\hline
\end{tabular}
\caption{Binding energies for various isoscalar $D_1 \bar{D}^*$ states in $L=0$ with $C=-$; 
the binding energies are given for a few values of $h$ in the range identified above.}
\label{table:isoscalarcharm}
\end{table}

If for example $h=1.0$, we find that two or even three S-states may arise, with binding energies 
$680\ \MeV$ (1S), $29\ \MeV$ (2S) and $7$ MeV (3S).  The rms radii, $r_{\rm rms}$ for these states are then 
approximately $0.2\ \fm$, $3\ \fm$ and $7\ \fm$ respectively.  This shows that the ground
state is typically hadronic, the 2S consistent with a canonical molecule and the 3S dubious.

Using Fig.~\ref{fig:FFEffect}, we can interpret the results for the $r_{\rm rms}$ values obtained for the S-wave 
states with the point particle potential.  The 1S state had an $r_{\rm rms}\approx 0.2$fm
clearly indicating that the state is bound in the first attractive well.  In contrast, the
2S state had an $r_{\rm rms}\approx 3$fm indicating that the particles are bound in the 
second attractive well of the potential.  The 3S state has an $r_{\rm rms}\approx7$fm
suggesting that it is bound by the third attractive well. 

If we take the potential, Eq.~\eqref{equ:swavepi_posspace}, and applied it
to $L>0$ systems unchanged apart from the centrifugal potential, we would find multiple bound states in the
P- and D-waves including some with binding energies of $\mathcal{O}(50)$MeV.  Firm conclusions
regarding the possiblity of such states would require a more extensive analysis of the origin of
Eq.~\eqref{swavepi}~than presented here.

The potential energy scales $\sim h^2$ but the binding energies are much more sensitive to $h$ 
(ground state binding energies scale like $\sim h^6$). This sensitivity to $h$ 
may not be unexpected, as the oscillatory nature of the potential in position space makes the potential 
turn over to repulsive when $r > 0.7$fm, and gives considerable sensitivity to these oscillations even for 
the short range 1S level, and critically so for the 2S. In a Coulomb potential the binding energies 
would scale as $\sim h^4$; this further explains the sensitivity noted above in the numerical calculation.

\subsection{Form Factor}
\label{sec:results:form-factor}

As noted previously, the form factor has a significant impact on the calculation of $h$ from the 
decay width.  
In the previous section we used this ``form-factor-renormalised" value of $h$, but otherwise continued to
treat the potential as if the hadrons were pointlike.
Therefore, it is prudent to investigate what effect form factors may have on the 
analysis of the molecular spectroscopy.  To examine this question we attach dipole form factors 
to the potential, Eq.~\eqref{swavepi}.  
Such ideas have been discussed in refs\cite{torn},\cite{newheavymesons} and \cite{liulambda}.
Following those ideas, we 
specifically choose to parametrise the form factor as
\begin{equation}
\mathcal{F} = \left( \frac{ \Lambda^2 - m_\pi^2}{\Lambda^2 - q^2 } \right)
\approx \left( \frac{ \Lambda^2 - m_\pi^2}{\Lambda^2 + \mu^2 - m_\pi^2 + \vec q^2 } \right)
\end{equation}
and the potential is multiplied by $\mathcal{F}^2$ -- we use the latter expression for ${\cal F}$.  
We have made the same static approximation as in Eq.~\eqref{swavepi}.
In nuclear physics dipole
form factors have only $\Lambda^2 + \vec q^2$ in the denominator due to the $\pi$ exchange between the 
nearly degenerate nucleons. 

\begin{figure}
  \includegraphics[scale=0.6]{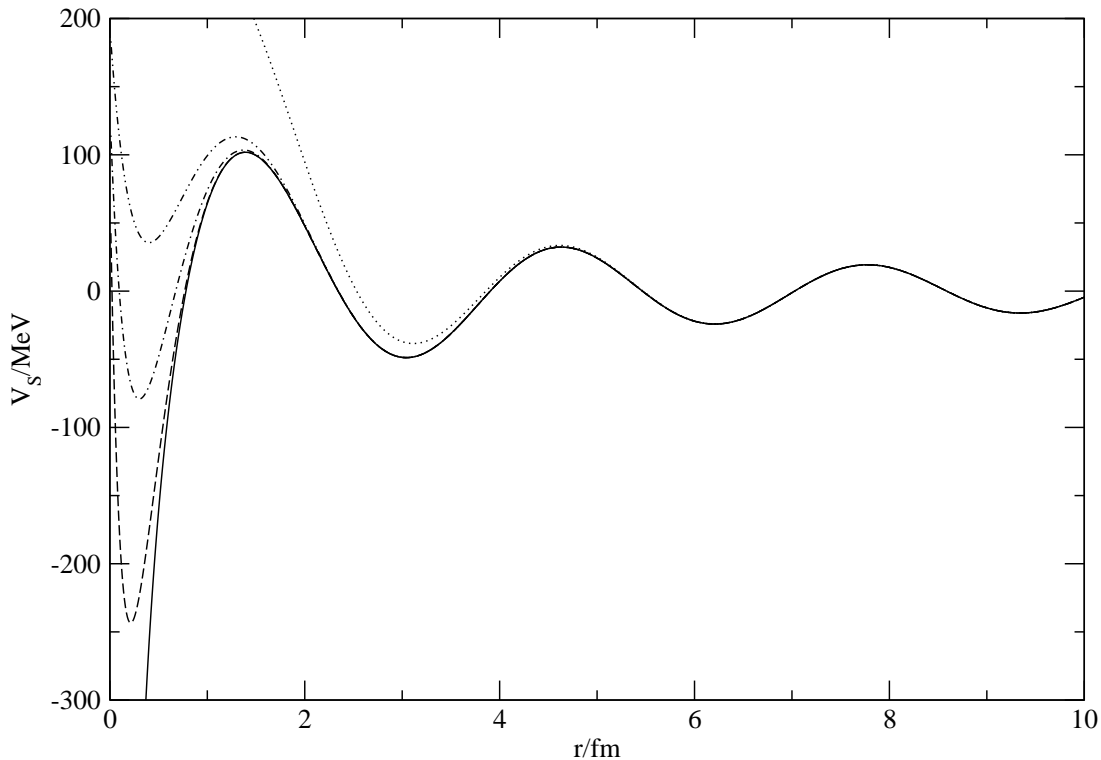}
  \caption{The potential, Eq.~\eqref{equ:swavepi_posspace_ff}, plotted against $r$ in the isoscalar $1^{--}$
  channel for $h=0.8$ and a variety of $\Lambda$s.  The solid line is the point particle result -- in effect
  $\Lambda \to \infty $; the dashed line 
  is the result for $\Lambda = 1.5$GeV; the dash-dot line is for $\Lambda = 1.0$ GeV; the dash-dot-dot line is for
  $\Lambda = .75$GeV; and the dotted line is for $\Lambda = .5$GeV.
  \label{fig:FFEffect}} 
\end{figure}

In position space, the form factor changes the potential from that in Eq.~\eqref{equ:swavepi_posspace} to:
\begin{equation}
V_S(r) = \frac{h^2 (m_{D_1}-m_{D^*})^2 }{8 \pi f_{\pi}^2} \left[ 
  \frac{\cos(\tmu r)}{r} - \frac{e^{-Xr}}{r} - \frac{(\Lambda^2 - m_{\pi}^2)}{2X} e^{-Xr} \right]
  \left( \tau_i \cdot \tau_j \right)
  \label{equ:swavepi_posspace_ff}
\end{equation}
with $X^2 \equiv \Lambda^2 + \mu^2 - m_{\pi^2} = \Lambda^2 - (m_{D_1}-m_{D^*})^2$ and 
where $\mu^2 \equiv -\tmu^2 < 0$.  We plot this potential for the isoscalar $1^{--}$ channel for various 
$\Lambda$ in Fig.~\ref{fig:FFEffect} and for a fixed $\Lambda$ and the various $1^-$ channels in 
Fig.~\ref{fig:PotPlots}.

Here $\Lambda$ is a purely phenomenological constant.  Although its value should be 
related to the convolution of the spatial wave functions of the hadrons, its value is fairly 
arbitrary in practice.  In the data-rich nucleon-nucleon sector, dipole form factors have been 
used in the Bonn nucleon-nucleon potentials.  In CD-Bonn one finds values of 
$\Lambda = 1.3 - 1.7 \GeV$\cite{cdbonn}.  However, there is no reason to believe that the value used 
in nuclear forces should be related to the value most appropriate for use in $\pi$ exchange 
between charmed mesons.  In the literature, other practitioners using dipole form factors in 
meson exchange molecular models employ values of: $\approx$1.2 GeV\cite{torn}, 
$\approx 1.2$-$2.3$GeV\cite{newheavymesons}, and $\approx 0.4$-$10$GeV\cite{liulambda}.

The qualitative effects of this form-factor are made apparent in Fig.~\ref{fig:FFEffect}.  
Regulating the potential introduces a soft-repulsive core instead of a singular attraction
at the origin.  As $\Lambda$ decreases, the first attractive well in the 
potential is entirely overwritten as a repulsive core.  

\begin{figure}[htp]
  \includegraphics[scale=.6]{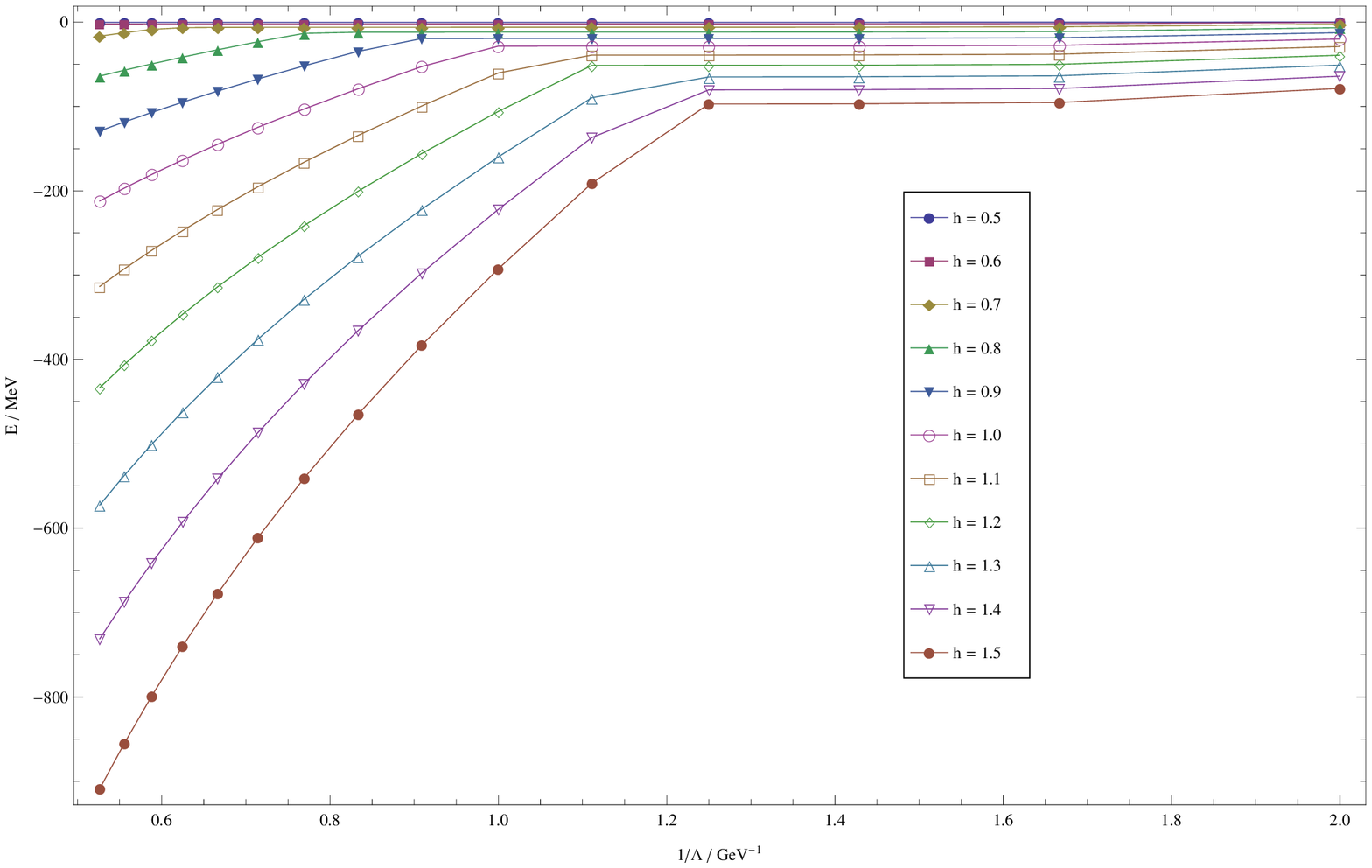}
  \caption{Plot of the 1S $1^{--}$ isoscalar binding energy for multiple values of $h$ as the form factor parameter
  $\Lambda$ is varied.}\label{fig:1SBE}
\end{figure}

\begin{figure}[htp]
  \includegraphics[scale=.6]{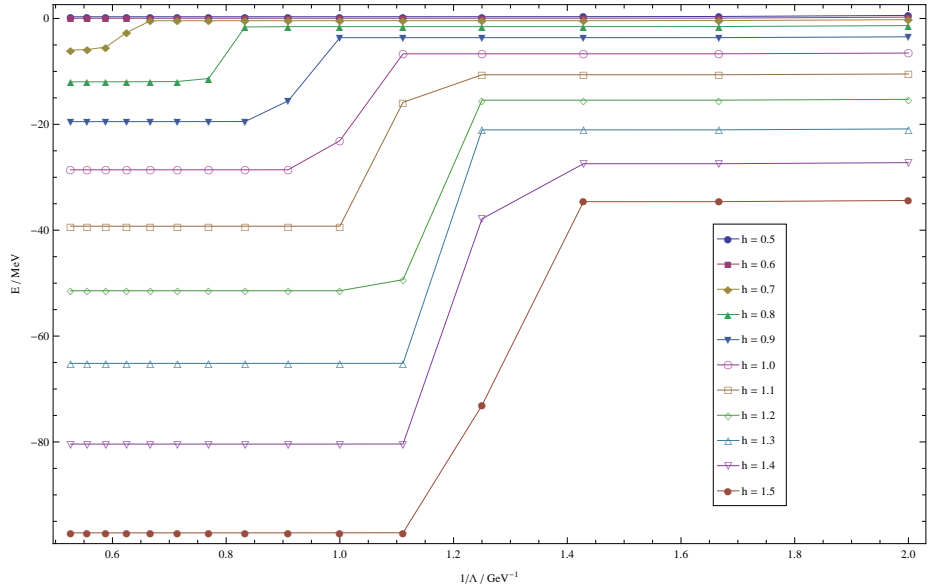}
  \caption{Plot of the 2S $1^{--}$ isoscalar binding energy for multiple values of $h$ as the form factor parameter
  $\Lambda$ is varied.}\label{fig:2SBE}
\end{figure}

We present the results for the binding energy as a function of $h$ and $\Lambda$ for the 1S 
and 2S isoscalar $1^{--}$ $D_1 \bar{D}^*$ states in Figs. \ref{fig:1SBE} and \ref{fig:2SBE}.  
The horizontal axes are $1/\Lambda$ so that the origin 
corresponds to the point-like case.  As one can see, the ground state binding energy 
falls off rapidly with decreasing $\Lambda$.
  Eventually the ground state binding energy finds a 
stable point and remains at approximately that energy for the rest of the considered values of 
$\Lambda$.  This behavior is sharply contrasted by the binding energy of the 2S state.  
The 2S binding energy is initially insensitive to a decrease in $\Lambda$ before 
falling steeply and finally becoming insensitive again.  

This behavior can be understood from the behavior of the potential in Fig.~\ref{fig:FFEffect}.
As $\Lambda$ is decreased from $\infty$, the potential is increasingly regulated.  This manifests
as overwriting the initial attraction from the potential and eventually replacing it by an entirely
repulsive core interaction for $\Lambda \lesssim 800$MeV.  Thus we would expect a steep fall off
in ground state binding energy as $\Lambda$ is decreased.  This expectation is borne out in 
Fig.~\ref{fig:1SBE}.  In contrast, the 2S state is bound primarily by the second attractive well,
 which is unaffected by decreasing $\Lambda$ as long as $\Lambda \gtrsim 800$MeV.  Thus, we
would expect the 2S binding energy to be relatively stable against decreasing $\Lambda$ as 
Fig.~\ref{fig:2SBE} confirms.  At some point,
which is $h$ dependent, there will no longer be enough attraction in the first attractive well
to bind the system, and so the ground state will begin to require presence in the second attractive well
in order to bind, displacing the 2S state and decreasing its binding energy.  When the 
first attractive well is completely overwritten, both the 1S and 2S states should have a relatively 
stable binding energy as the attractive wells (second and third) which bind them are stable against
decreased $\Lambda$.  Indeed the binding energies of the 1S state decrease slightly as $\Lambda\to 500$MeV
corresponding with the alteration of the second attractive well in Fig.~\ref{fig:FFEffect}.


This analysis shows that the molecular spectroscopy is very sensitive to the parameters. 
While a simple abstraction of parameters from existing data support the idea that a 
spectroscopy of molecules could arise, one cannot with certainty predict this.
However, the result of strong binding appears relatively robust.  Indeed our results 
show that the existence of robust bound states
(assuming $h$ is sufficently large) does not depend on deep attraction at the origin, and
that, even in the presence of a strong repulsive core interaction, a bound state
should exist with a binding energy largely determined by long-range 
($\gtrsim 2$fm) virtual pion effects.

\begin{figure}[htp]
  \includegraphics[scale=0.6,angle=-90]{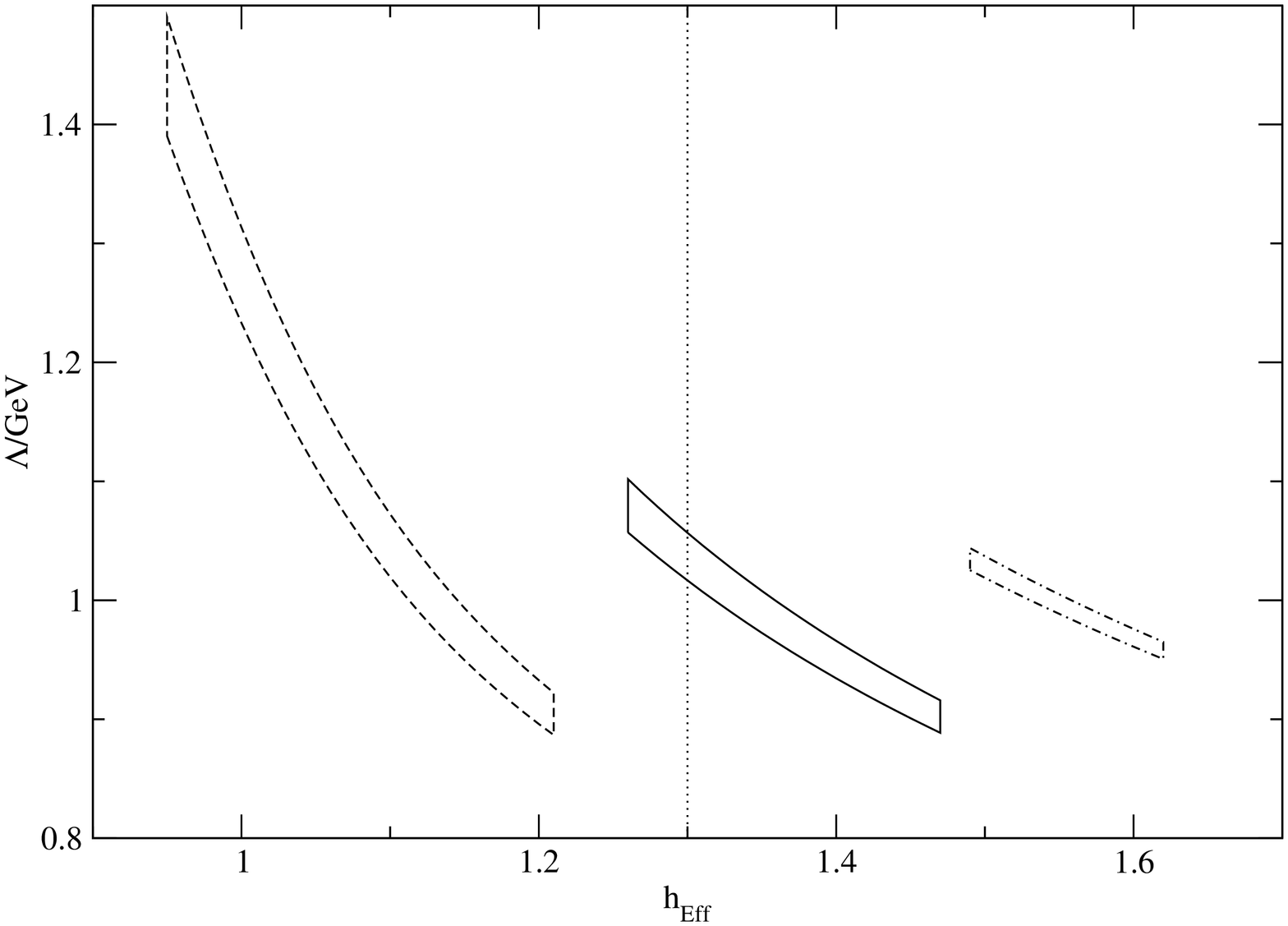}
  \caption{Countour plot of the values of $h_{\rm Eff}$ and $\Lambda$.  The interior of the boxes corresponds to values
  which reproduce the binding energies of the $Y(4260)$ and $Y(4360)$ to within errors.  The center
  box is for the experimental $D_1$ mass while the box on the left is for the $D_1$ mass minus its
  error and the box on the right is the $D_1$ mass plus the error.
  The dotted line 
  corresponds to the value for $h_{\rm Eff}=1.0+0.3$ from the $D_1$ experimental width.\label{fig:YsFit}}
\end{figure}

If the $Y(4260)$ and $Y(4360)$ are confirmed as genuine signals, then within this simple modelling,
their energies are qualitatively consistent with those expected for S-wave binding. 
Indeed, the differing sensitivities of the 1S and 2S states to $\Lambda$ would 
allow one to tune the model to reproduce the binding energy of the $Y(4260)$, $174\pm9$ MeV
and the $Y(4360)$, $76\pm 13$ MeV assuming the mass of the $D_1$ was exactly 2427 MeV.  Since 
the mass of the $D_1$ affects both binding energies in a systematic way, we cannot simply add its 
error in quadrature for both to obtain our binding energies with their error.  Instead,
we study the system for $m_{D_1} = 2427-40$MeV; $2427$MeV; $2427 + 40$ MeV requiring binding energies of:
$134\pm 9$MeV, $36\pm13$MeV; 
$174\pm 9$MeV, $76\pm13$MeV; 
$214\pm 9$MeV, $116\pm13$MeV.
We present the ``tune-ability'' of the model in Fig. \ref{fig:YsFit}.

The mass of the $D_1$ effects the potential in two straightforward ways.  First, it 
factors into the calculation of $h$ from the decay widths.  Secondly, it helps determine
the mass gap which, along with $h$ controls the strength of the potential.  Although the 
value of $h$ and the mass gap depend on the value of the $D_1$ mass, 
\begin{equation}
  h_{\rm Eff} = h\frac{m_{D_1} - m_{D^*}}{2427 - m_{D^*}}
\end{equation}
is unchanged as $m_{D_1}$ varies over its error.  This allows us to plot the different mass
cases on a single axis.  (The mass of the $D^*$ has an insignificant error.)

Fig. \ref{fig:YsFit} was produced by parameterizing the binding energies.  We assumed that
the 2S binding energy was approximately independent of $\Lambda$ and so could be used
to determine $h$.  
This assumption has been explicitly verified for the values of $h$, $\Lambda$ considered
and is found to be a good approximation.
Then the 1S binding energies were parameterized as
a quadratic function of $1/\Lambda$ whose coeffecients were quadratic functions of $h$.
This parameterization reproduced the 1S binding energies over the relevant range
of these parameters.  The quadratic formula could then be used to extract the range of
$\Lambda$ from the $Y(4260)$ binding energy at each $h$.  

The region of compatability extends to just below the error bounds for $h_{\rm eff}$ to slightly
above it.  $\Lambda$ values are undetermined by experiment, however the compatible values 
lie around 1 GeV which is near values used by other practictioners.
Thus, a consistent, physically reasonable parameterization
of $h$ and $\Lambda$ is possible which permits the identification of the 
$Y(4260)$ and $Y(4360)$ as the 1S and 2S bound states of the $D_1 \bar{D}^*$ system 
respectively.  

In general within the chiral model, 
$h$ is a function of the experimental  $\Gamma (D_1 \to D^* \pi)$.
  If experiment were to show that the width were different than the current values that we 
have used,
the consequent alteration of the molecular binding energies could be considerable. It is here that some 
major limitations in the robustness of the model lie.

\subsection{Other $D_1$ $D^*$ bound states and flavour exotics}
\label{sec:results:othercharm}

The potentials in the $C=\pm$ and isovector/isoscalar channels are related by a simple constant.
The potentials of isovector and isoscalar channels are related by a $\tau_i\cdot\tau_j$ factor 
while the potential in channels with opposite charge conjugation are related by a relative phase.
Therefore, we can use Fig.~\ref{fig:FFEffect} to interpet the binding in
all these channels against finite size effects and a possible repulsive core.  The robustness
of our results in the isoscalar $1^{--}$ channel were discussed previously.

\begin{figure}
  \includegraphics[scale=0.6]{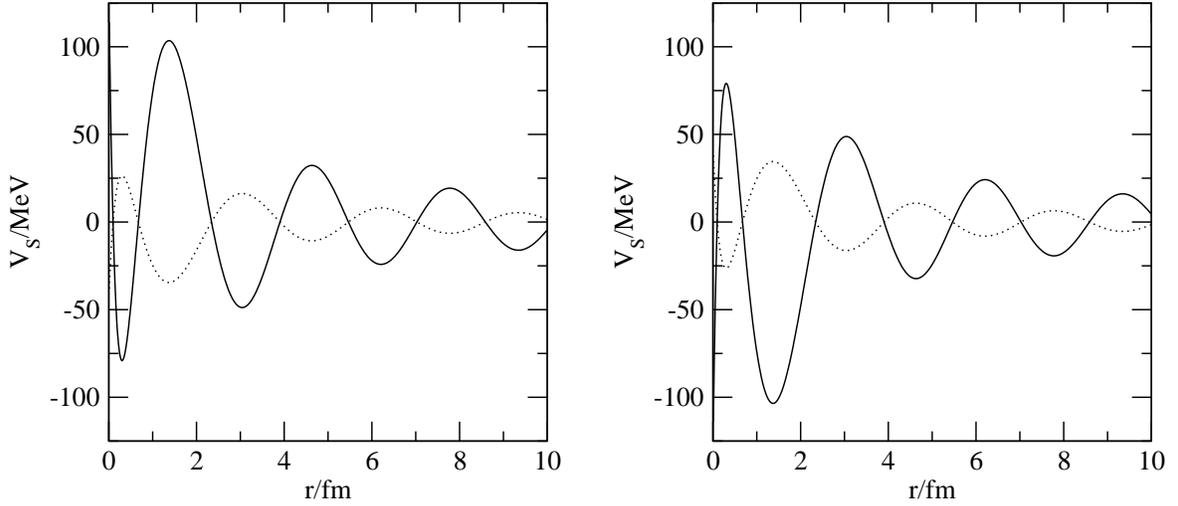}
  \caption{The potential, Eq.~\eqref{equ:swavepi_posspace_ff}, plotted against $r$ in the various $1^-$ 
  channels for $h=1$ and $\Lambda = 1$GeV.  The dotted lines are the isovector potentials while the 
  solid lines are the isoscalar potentials. The left panel shows the $C=-$ potentials and the right 
  panel gives the $C=+$ potentials.
  \label{fig:PotPlots}
  }
\end{figure}

In Table \ref{table:morecharm} we show the binding energies of some of the low-lying isoscalar 
and isovector $D_1 \bar{D}^*$ states in relative S-wave with $C=\pm$.  
Binding energies are given for a few values of $h \in [0.8, 1.3]$ and $\Lambda=1$ GeV.
Interestingly we find potentially robust binding in all isospin and charge-conjugation states.

\begin{table}[htp]
\begin{tabular}{l c|c|c|c|c|c|c|}
& & \multicolumn{6}{c}{Binding Energy / MeV} \\
  State & Isospin      & $h=0.8$ & $h=0.9$ & $h=1.0$ & $h=1.1$ & $h=1.2$ & $h=1.3$ \\
\hline
1S $(0,1,2)^{--}$  & 0 & $12$     & $20$    & $29$    & $60$      & $110$  & $160$   \\
2S                 &   & $1.6$    & $3.6$   & $23$  & $39$    & $51$ & $65$  \\
3S                 &   & --       & $0.7$   & $6.7$   & $11$    & $15$ & $21$  \\
\hline 
1S $(0,1,2)^{--}$  & 1 & $4.2$  & $8.8$  & $15$ & $21$ & $29$ & $38$ \\
2S                 &   & $-$  & $-$  & $0.2$  & $0.7$  & $1.5$  & $2.8$  \\
\hline 
1S $(0,1,2)^{-+}$  & 0 & $47$ & $67$ & $90$ & $120$  & $150$  & $180$  \\
2S                 &   & $4.2$  & $8.1$  & $13$ & $19$ & $27$ & $35$ \\
3S                 &   & $0.5$  & $1.6$  & $3.5$  & $6.1$  & $10$  & $14$ \\
\hline
1S $(0,1,2)^{-+}$  & 1 & $0.1$ & $0.5$ & $1.6$ & $3.4$ & $5.9$ & $8.9$ \\
2S                 &   & $-$   & $-$   & $-$   & $0.1$ & $0.4$ & $0.9$ \\
\hline
\end{tabular}
\caption{Binding energies for $D_1 \bar{D}^*$ states with various isospins and charge conjugations; the binding energies 
are given for a few values of $h$ in the range identified above and $\Lambda = 1$GeV.}
\label{table:morecharm}
\end{table}

We note that the pattern of binding described here is valid for $\Lambda=1$ GeV and the pattern
will be altered as $\Lambda$ changes.  In particular, the pattern will change
if the finite size effects wipe away less of the deep attractive core which binds the isoscalar $1^{--}$ 
and isovector $1^{-+}$ channels.  In general a higher value of $\Lambda$ will lead to (significantly) more deeply 
bound isoscalar $1^{--}$ and isovector $1^{-+}$ bound states, and slightly less bound isovector $1^{--}$
and isoscalar $1^{-+}$ states. 
  
The pattern of relative binding energies between the channels may be understood from
Fig.~\ref{fig:PotPlots}.  The most deeply bound states occur in the isoscalar $1^{-+}$ channel
where the potential is repulsive near the origin but has a deep attractive well (due again to
the isospin factor of 3) near 1fm.
The second most deeply bound states occur in the isoscalar $1^{--}$ channel where 
the $\tau_i\cdot \tau_j$ term contributes a --3 factor and there is a deep attraction near 
the origin.  We can see in Fig. \ref{fig:PotPlots} that the form-factor has reduced 
 the magnitude of the first dip in the oscillating potential (around 0.3 fm), 
 making it smaller than the first bump (around 1.2 fm).  
This is why the isoscalar $1^{-+}$ channel has deeper binding than the $1^{--}$ channel.  
The isovector channels lose the isospin factor of 3, leading to significantly reduced binding
in these channels.  However, their relative binding is the same: the $1^{--}$ potential retains the deeper 
attraction near 1 fm whereas the isovector $1^{-+}$  channel loses the attraction around the origin
due to form-factor effects.  Hence the isovector $1^{-+}$ channel is the least deeply bound when $\Lambda =1$ GeV.
Consequently the prediction of bound states in the isovector $1^{-+}$ is the least robust.  

The situation is very different for the isovector $1^{--}$ and isoscalar $1^{-+}$ channels.
In both of these channels the point particle potential is repulsive at the origin and they must
bind in the first attractive well which is $\approx$1fm away from the origin.  Therefore we expect these
numerical results to be robust against finite size effects and a repulsive core.  However,
in the presence of intense regulation of the potential then, in this channel,
the deep attraction being overwritten 
to strong repulsion with decreasing $\Lambda$, shown in Fig.~\ref{fig:FFEffect}, becomes a strong
repulsion being overwritten to deep attraction.  Therefore, we conclude that the existence of 
deep binding in these channels is a very robust result which should be insensitive to 
strong, short-range dynamics and totally independent of finite size effects of the potential, 
though both may contribute to deeper binding.

The ranges of $h$ and $\Lambda$ which reproduce binding energies for the $Y(4260)$ and $Y(4360)$ 
(see Fig. \ref{fig:YsFit}) are of particular interest.  
The case h=1.3 (Table \ref{table:morecharm}) illustrates how it is possible to identify the 
1S and 2S 1$^{--}$ respectively as $Y$(4260) 
and $Y$(4360). In such a scenario it is possible that a third 1$^{--}$ state could occur around 
4400 MeV. 
But of most interest is the prediction of a robust isoscalar exotic 1$^{-+}$ bound state in the 
vicinity of, or even below, the $Y$(4260). 
If this exotic state is below the $Y$(4260) then it may possibly be observed through 
$Y(4260)\rightarrow Y(4200?) +\gamma$.

Table \ref{table:morecharm} shows binding in both isovector $1^-$ channels.
We therefore must reverse our previous concurrence\cite{cdprl} with the 
conclusions of Ref.~\cite{liu}: when subjected to a more complete analysis we find that a bound 
state may exist due to one pion exchange between $D_1\overline{D^*}$ in the isovector $1^-$ channel.

We find it interesting to note that the $Z(4430)$ has a mass of $4433 \pm 4$ MeV\cite{z4430}.  Therefore if it were
a $D_1\overline{D^*}$ molecule, it would have a binding energy of $4\pm 9$ MeV.  This binding energy is 
compatible with a charged partner of the $1^{-+}$ isovector result for the range of $h$ and $\Lambda$ which reproduces the 
$Y(4260)$ and $Y(4360)$.  
A more complete analysis than that provided here is necessary to make a definitive identification.  However
we find the possiblity that one pion exchange might provide a consistent description of the $Y(4260)$, $Y(4360)$,
and the $Z(4430)$ with physically reasonable parameters encouraging.

In addition, we predict doubly charmed ($D_1 D^*$ as opposed to $D_1 \overline{D^*}$) isoscalar and 
isovector states degenerate with respectively the isoscalar and isovector $D_1 \overline{D^*}$ states in $C=-$.  
We refer to Ref.\ \cite{fcthomas} for a discussion of the signs involved.

\subsection{Bottom analogues}
\label{sec:results:bottomkaon}

In Table \ref{table:isoscalarbottomkaon} we present the binding energies of some of the low-lying isoscalar 
$B_1 \bar{B}^*$ states in relative S-wave with $C=-$, along with the analogous 
$D_1 \bar{D}^*$ states for comparison.  Binding energies are given for a few values of $h \in [0.8, 1.3]$ and
$\Lambda = 1$ GeV.

\begin{table}[htp]
\begin{tabular}{l|c|c|c|c|c|c|}
 & \multicolumn{6}{c}{Binding Energy / MeV} \\
  State                   & $h=0.8$ & $h=0.9$ & $h=1.0$ & $h=1.1$ & $h=1.2$ & $h=1.3$ \\
\hline
1S $D_1 \bar{D}^*$ $(0,1,2)^{--}$  & $12$   & $20$  & $29$  & $60$    & $110$  & $160$   \\
2S                                 & $1.6$    & $3.6$   & $23$  & $39$    & $51$ & $65$  \\
3S                                 & --       & $0.7$   & $6.7$   & $11$    & $15$ & $21$  \\
\hline
1S $B_1 \bar{B}^*$ $(0,1,2)^{--}$  & $56$  & $93$  & $140$ & $190$  & $250$ & $320$ \\
2S                                 & $20$  & $29$  & $38$  & $49$   & $61$  & $75$  \\
3S                                 & $6.3$ & $9.8$ & $14$  & $19$   & $24$  & $30$ \\ 
\hline 
\end{tabular}
\caption{Binding energies for various isoscalar $B_1 \bar{B}^*$ and $D_1 \bar{D}^*$ states in $L=0$ with $C=-$; 
the binding energies are given for a few values of $h$ in the range identified above.}
\label{table:isoscalarbottomkaon}
\end{table}

The binding energies are generally deeper than in the charmed analogues.
This is easily understood: the higher mass of the $B$ mesons 
result in a lower kinetic energy.  In general we predict analogous effects in the $B$ analogs of the charmed
system, subject to differences in the width which is experimentally undetermined for the $B_1$.

Similar effects may exist in the $K$ system.  However the phenomenology of the $K_1$ is more complex
and the heavy quark approximation is certainly inadequate.  Together with
the constraint on $h$ implied by the width, this prevents us making quantitative conclusions, 
we only note the qualitative possibility that S-wave pion exchange may produce binding in the $K$ system.

\section{Discussion}
\label{sec:Uncertaintities}

The results for binding energies, and even whether states bind at all, are sensitive to parameters, and also to
more complicated (possibly more realistic) modelling of the strong interactions.

We have focussed solely on the $t$-channel force from virtual pion
exchange, specifically, the four fermion intermediate states in the Fock state. 
Therefore, we have taken only the real part of the potential and solved the Schr\"{o}dinger 
equation and ignored the imaginary part arising from the exchange of a real, on-shell, pion.
There are also $s$-channel forces arising from intermediate $c\bar{c}$ excited states.
More immediately in our molecular approach there are intermediate states with a real pion, of form $D^*\pi \bar{D^*}$.

The ability of a virtual exchanged particle to be on-shell introduces an imaginary component 
to the matrix element and, hence, to the potential.
The effect  is to make the energy complex: 
the real part is taken as the binding energy while the imaginary part 
is interpreted as the width of the state. That the on-shell intermediate state 
would manifest as a width seems natural as it represents a direct connection 
between the bound state and a possible decay channel.

The picture is then that the $D_1$ decays into a $D^*\pi$ and the ``would be" 
quasi-molecular bound state disintegrates, or even fails to form. 
Thus, we expect the on-shell pion contribution will endow any state 
produced by this mechanism with a width, or that it produces a non-resonant background which may obscure the signal.

Within our approximations we find deeply bound meta-stable states.
The $D^*\pi \bar{D^*}$ generates widths and background. Whether these states 
remain visible is then dependent upon the relative importance of
neglected forces, such as mixing with $c\bar{c}$ or $D^*\pi \bar{D^*}$. 
In general it is difficult to calculate 
the impact of neglected effects, not least because strong interactions
are complicated and we are approximating one particular force as dominant.

If the $Y(4260)$ is an example of our states, then its visibility shows that 
Nature is kind, at least in the $1^{--}$ channel. It has given a width
of ${\cal O}(100)$MeV and a visibility above background. It could be that this 
fortune is because a $c\bar{c}$ component drives the production, and the $D_1\bar{D^*}$
rearrangement then drives the $\psi \pi\pi$ signal.

Thus the conclusion of this analysis is that while it is {\it possible} that a deep bound molecular spectroscopy
with signals visible above a background 
can arise, {\it it is
not mandatory}. However, as we have already noted,
the appearance of $Y(4260)$ and $Y(4360)$ are consistent with being the first two states observed in 
such a spectroscopy. The immediate test of this is to seek evidence for these states in $D\bar{D}\pi\pi\pi$. Unless there is
some dynamical suppression, such channels must show strength if a $D^*\bar{D_1}$ bound system is present. If this first test is passed, then
a search for other transitions and evidence for analogous states in $B^*\bar{B_1}$ would be warranted. In this latter case we note the 
apparent presence of anomalous state $\Upsilon(10.88)$\cite{georgehou}

This and other phenomenological implications are the theme of the next section.

\section{Phenomenology}
\label{sec:Phenomenlogy}
We have studied the $D^*\bar{D_1}$ molecules and found deeply bound states with $I=0$, 
which are degenerate for the $(0,1,2)^{--}$ channels. 
However, the number of
 potentially deeply bound states is very sensitive to parameters. Typically we anticipate the
binding energies of the $I=0$ states to have the
orders of magnitude as follows:
 1S $O(10-100)$MeV; 2S $O(1-10)$MeV, with an exotic $1^{-+}$ between the
1S and 2S levels.

Further reasons to anticipate a rich spectroscopy are that this S-wave $\pi$  
exchange also can occur for $D_0\to D$ and the off-diagonal
$D\bar{D_1} \to D_0\bar{D^*}$. The strengths for each of these in the heavy quark limit are identical. In practice 
there will be model dependent perturbations due to mass shifts and mixings; these are beyond 
the present paper and only merit study if the general features of our model show up in the data.

In general: 
if S-wave pion exchange forms
deeply bound charmed molecules comprised of $D^*\bar{D_1}; D^*\bar{D_0};D\bar{D_1};D\bar{D_0}$ (and manifest charm analogs)
, there will be a rich spectroscopy of states in the
$3.9 - 4.5$ GeV mass range. These can include states that are superficially charmonium, such as $I=0$, $0^{-+}$ 
and $1^{--}$,
as well as exotic $J^{PC}: 0^{--} $ and $1^{-+}$. In addition there are also
 states with charmonium character but $I=1$. 
States such as 
$I=0$, $0^{-+}$ and $1^{--}$ may contain \cc~ in their Fock state and hence be produced at measurable rates; 
the other states
have no such aid, but may be produced in radiative or strong transitions from higher lying molecules.
Manifestly charm ($D^*D_1$, etc.) states are also expected and are degenerate with the $C=-$ charmonium like 
states.
The pattern and
observability of these will depend on the detailed pattern of the spectroscopy.

    The states that are most amenable to experimental study are the $I=0$, $1^{--}$. These
    occur in $D_1D^*$, and also can arise from the off-diagonal
   S-wave potential for $DD_1 \to D_0 D^*$. Hence there can be a rather rich spectroscopy in the $1^{--}$ sector.
   As there are apparently several states of varying statistical
   significance emerging in the data, we shall primarily focus on this channel here.
   
The best established enigmatic structure in the $1^{--}$ sector here is
$Y(4260)$ which is seen in $\psi \pi\pi$. Its typical hadronic width $\Gamma(4260) \sim 90$MeV 
implies that either $\psi \pi\pi$ is not the dominant decay channel or that 40 years
of experience with the OZI rule and strong interactions is wrong. Given the nearness of the $D(L=0)+D(L=1)$ thresholds
which can be accessed in S-wave, rearrangement into $\psi \pi\pi$ at low momentum seems reasonable, and has been
invoked as a qualitative explanation of these phenomena\cite{closepage}.

As $D^*$ and $D_0 \to D\pi$, whereas $D_1 \to D \pi\pi$, 
then if the dynamics are  associated with the nearby $D\bar{D_1}$ and $D^*\bar{D_0}$ thresholds,
such as $Y(4260)$ being a $D\bar{D_1}$ molecule, or a hybrid \cc~ that is dynamically attracted
towards that threshold, then
strength should be seen in the $D \bar{D} \pi\pi$ channels\cite{closepage,closetalk}.
However, if the $Y(4260)$ is a $D^*\bar{D_1}$ bound state, then the favored strong decay will be 
$\to D\bar{D} 3 \pi$ in contrast to the aforementioned $D\bar{D_1}$ 
or $D^*\bar{D_0} \to D\bar{D} 2\pi$.

A preliminary report from Belle\cite{belle2pi} sees no evidence for $D^*D\pi$ in the $Y(4260)$
region. This disfavors $D\bar{D_1}$ and potentially also the $D\bar{D}\pi\pi$ channel. Thus by
default, the possibility that the strength is driven by $D\bar{D}\pi\pi\pi$ becomes tantalizing.

Thus an immediate consequence 
of this interpretation is that if the $Y(4260)$ is a $D^*\bar{D_1}$ molecule,
there {\it must} be significant coupling of the $Y(4260) \to 
D\bar{D}\pi\pi\pi$
that could exceed that to $D\bar{D}\pi\pi$.
More generally an unavoidable conclusion of this dynamics is that in the $1^{--}$ sector 
the $e^+e^- \to D \bar{D} \pi\pi\pi$ channel 
has significant strength in the region of any $D^*\bar{D_1}$ molecular states.
Hence we urge measurement of the relative importance of the channels $e^+e^- \to D \bar{D} \pi\pi\pi$
and of $e^+e^- \to D \bar{D} \pi\pi$ (when, in the latter, $D^*\bar{D^*}$ has been removed).

The depth of binding of the ground state with trial wave functions  already suggested\cite{cdprl} the
tantalizing possibility that a radially
excited state could also be bound. Numerical solutions of the Schrodinger equation confirmed that this
is likely to be the case in the range of models discussed here.

The excitation energy for radial excitation of a compact
QCD \cc~state is $\mathcal{O}(500)$MeV;  it takes less energy $\mathcal{O}(100)$MeV
to excite the extended molecular system which has no linearly
rising potential. The spatial extent of the molecular $2$S system is significantly greater
than that of \cc~  hadrons. The rearrangement of constituents leading to
final states of the form $\psi$ + light mesons then rather naturally suggests that the lower (radial) 
states convert to $\psi \pi\pi$
( $\psi' \pi\pi$) respectively. In this context it is intriguing that there are states observed with
energies and final states that appear to be consistent with this:
$Y(4260) \to \psi \pi\pi$ \cite{4260} and the possible higher state $Y(4360) \to \psi' \pi\pi$
\cite{babar1,belle1} are respectively 170MeV and 70 MeV below the
$D^*(2010)\bar{D_1}(2420)$ combined masses of 4430MeV. Here again, for a $D^*\bar{D_1}$ molecular state, we would
expect significant coupling to $D\bar{D}\pi\pi\pi$.

If these states were to be established as members of molecular systems, one could tune the model accordingly. 
Further, this 
could be an interesting signal for a $D^*\bar{D_1}$ quasi-molecular spectroscopy with transitions
among states that could be revealed in, for example, $e^+e^- \to \psi\gamma\gamma\pi\pi$.
Indeed, if we identify $1$S$(4260)$ and $2$S$(4360)$, then we expect the exotic $1^{-+}$ to occur in the
vicinity of the $Y(4260)$.
Given that lattice QCD finds activity for a hybrid cc* signal in this channel in this region, 
one should now actively search for evidence. A clear signature is that
the $1^{-+}$ hybrid will couple to $D\bar{D}\pi\pi$ in either the $D^*\bar{D_0}$ or $D\bar{D_1}$ combinations; 
looking for the presence of strength in $e^+e^- \to \gamma D\bar{D}\pi\pi$
which does not include $D^*\bar{D^*}$ should thus be a primary endeavor. 
The absence of such a channel could have far
reaching implications for theory. 

   While our discussion has centered on charmonium, the remarks hold more generally.
   Since the attraction of the potential depends only on the quantum numbers of the light $q\bar q$, 
it follows immediately that the flavor of the heavy quarks is irrelevant, at least qualitatively. 
Hence we expect similar effects to occur in the $b\bar{b}$ and $s\bar{s}$ sectors. 
It has been noted that $\Upsilon(10.86)$GeV
appears to have an anomalous affinity for $\Upsilon \pi\pi$\cite{upsilon}.  This state is $\sim 130$
MeV below $B^*\overline{B}_1$ threshold. In the $\phi\pi\pi$ channel there is an enhancement at 
2175MeV\cite{phi}.  
This is approximately 125MeV below the $K^*\overline K_1(1400)$ threshold.  This is
consistent with the $K^*\bar{K_1}$ spectroscopy; however, as commented earlier, analysis here is 
less reliable, 
as  the heavy quark approximation fails, and the phenomenology of the $K_1(1270;1400)$
pair is more complicated\cite{barnes,lipkin}.

The primary test for this picture is that if the states in the $4$ to 4.5 GeV region are deeply bound $D^*\bar{D_1}$
spectroscopy, then their decays in charm pairs must show strength in the $D\bar{D}\pi\pi\pi$ channels. The
energy dependence of this channel and that of $D\bar{D}\pi\pi$ (with no $D^*\bar{D^*}$) can reveal the
mixings between $D^*\bar{D_1}$ and  $D\bar{D_1}/D^*\bar{D_0}$ molecular systems. The presence of exotic 
$1^{-+}$ is also expected.

\section{Conclusions}
\label{sec:Conclusions}

In general we find that deeply bound molecules in the $D_1\bar{D^*}$ system should occur
as a result of $\pi$ exchange in S-wave, 
leading to a potentially rich spectroscopy. 
Whether such states are narrow enough to
show up above background is a question that experiment may resolve. We note however that
the emerging data on the $1^{--}$ states known as $Y(4260)$ and $Y(4360)$ are consistent with
being examples of these molecular states. The immediate test is to verify if the 
prominent channels with manifest charm in this mass region are $D\bar{D}\pi\pi\pi$. If this is confirmed,
then more detailed studies will be merited,
in particular searches for an exotic $1^{-+}$ in the vicinity of 4.2GeV. This state
could be produced via $Y(4260) \to (1^{-+}) + \gamma$, and/or be revealed in
1$^{-+} \to \psi \gamma$.

Table \ref{table:morecharm} with h=1.3 shows a possible spectroscopy consistent with the
Y(4260) and Y(4360) as the 1S and 2S $1^{--}$ states. In this case, the
exotic states expected are I=0 $0^{--}$ also at 4260 and 4360 
(in both charmonium-like and manifestly charm channels);
the isoscalar $1^{-+}$ at 4250 and 4395; and also I=1 ``charmonium" states,
including $1^{--}$ at 4390.

As long as one picks and chooses which datum one will fit, it is possible to fit it in a molecular model.
A reason is that binding energies are very sensitive to parameters that are not well determined elsewhere. Thus a model designed to fit a single state has limited appeal.
The more relevant test is whether a group of states share a common heritage, and their production or decay properties
reveal the underlying molecular structure.
In the particular case here, one can fit the masses and decay widths in tetraquark, hybrid and molecular models.
As such the existence of these states does not discriminate among them.

However, the pattern of $J^{PC}$ and the decay channels differ. Thus the sharpest tests of their dynamical structure
appears to be in the decay branching ratios. Hence, for example, the $Y(4260)$ as a $cs\bar{c}\bar{s}$ tetraquark would couple to
$D_s\bar{D_s}$; a hybrid or molecule associated with $D\bar{D_1}$ threshold would be expected to appear in $D\bar{D_1} \to D\bar{D}\pi\pi$;
molecules associated with the $D^*\bar{D_1}$ threshold by contrast would have significant strength in the $D\bar{D}\pi\pi\pi$ channels.
Thus the decay branching ratios of states seem likely to be sharper indicators of their dynamical nature than simply
their masses.

If our hypothesis is correct, we expect significant strength in the $e^+e^- \to D\bar{D}\pi\pi\pi$
channels in the 4 to 5 GeV region. Such evidence may already exist among the data sets
for $e^+e^-$ annihilation involving ISR at BaBar and Belle.

\section*{Acknowledgements}

One of us (FEC) thanks Jo Dudek for a question at a Jefferson Lab seminar which stimulated some of 
this work and T. Burns for discussion.  This work is supported by grants from the Science \& 
Technology Facilities Council (UK), in part by the EU Contract No. MRTN-CT-2006-035482, ``FLAVIAnet.', 
and in part authored by Jefferson Science Associates, LLC under U.S. DOE Contract No. DE-AC05-06OR23177. 
The U.S. Government retains a non-exclusive, paid-up, irrevocable, world-wide license to publish or 
reproduce this manuscript for U.S. Government purposes.


\end{document}